# An Aspect-Oriented Approach for SaaS Application Customization

Ashraf A. Shahin, Areeg Samir, and Abdelaziz Khamis

Department of Computer and Information Sciences,
Institute of Statistical Studies & Research, Cairo University, Egypt

## Abstract

Multi-tenancy is one of the most important concepts for any Software as a Service (SaaS) application. Multi-tenant SaaS application serves a large number of tenants with one single application instance. Complex SaaS application that serves significant number of tenants could have a huge number of customizations with complicated relationships, which increases the customization complexity and reduces the customization understandability. Modeling such customizations, validating each tenant's customization, and adapting SaaS applications on the fly based on each tenant's requirements become very complex tasks. To mitigate these challenges, we propose an aspect-oriented approach that makes use of the Orthogonal Variability Model (OVM) and Metagraphs. The OVM is used to provide the tenants with simple and understandable customization model. A Metagraph-based algorithm has been developed to validate tenants' customizations. On the other hand, the aspect-oriented approach offers a high level of runtime adaptability.

**Keywords:** Cloud computing, Software as a Service (SaaS), SaaS Application Customization, Aspect-Oriented Programming, Orthogonal Variability Model, Metagraph, AO4BPEL.

## 1. Introduction

Cloud computing is a model for enabling convenient, on-demand network access to a shared pool of configurable computing resources that can be rapidly delivered with a minimal management effort or service provider interaction [1].

Software as a Service (SaaS) is a software delivery model in which software resources are accessed remotely by clients [2]. The SaaS delivery model is focused on bringing down the cost by offering the same instance of an application to as many customers, i.e. supporting multi-tenants. Multi-tenancy is one of the most important concepts for any SaaS application.

SaaS applications have to be customizable to fulfill the varying functional and quality requirements of individual tenants [3]. The elements of an application that need to be customized include: Graphical User Interface (GUI), workflow (business process logic), service selection and configuration, and data [4]. Several researches have attempted to support customization of these elements [2, 5, 6, 7, 8].

One of the most prominent approaches for achieving the tenant-specific customization is providing an application template with unspecified parts that can be customized by selecting components from a provided set of components [8, 9]. These unspecified parts are called customization points of an application [10].

SaaS application customization research literature points out that there are four key concerns that need to be addressed: first, how to model the customization points and variations, second, how to describe the relationships among variations, third, how to validate customizations performed by tenants, and fourth, how to associate and disassociate variations to/from customization points during run-time [11].

Based on Orthogonal Variability Modeling (OVM) and Metagraphs, we propose an aspect-oriented approach for customizing SaaS applications which addresses the above concerns. OVM is used to model customization points and variations, and to describe the relationships among variations. A Metagraph-based algorithm has been developed to validate tenants' customizations. And finally, an aspect-oriented extension for the Business Process Execution Language (AO4BPEL) [12, 13] is used to associate and disassociate variations to/from customization points during run-time.

The remainder of the paper is organized as follows. In section 2, we give some background material on OVM, Metagraphs and AO4BPEL. Section 3 gives a short overview of related work. Section 4 presents our proposed approach to customize SaaS applications. In section 5, we evaluate the proposed approach. Finally, we conclude in section 6.

## 2. Background

### 2.1 Orthogonal Variability Modeling (OVM)

OVM is a proposal for documenting software product line variability [11]. In OVM only the variability of the product line is documented. In this model a *variation point* (*VP*) documents a variable item and a *variant* (*V*) documents the possible instances of a variable item. All VPs are related to at least one V and each V is related to at least one VP. Figure 1, taken from [11], shows for the graphical notation used in OVM.

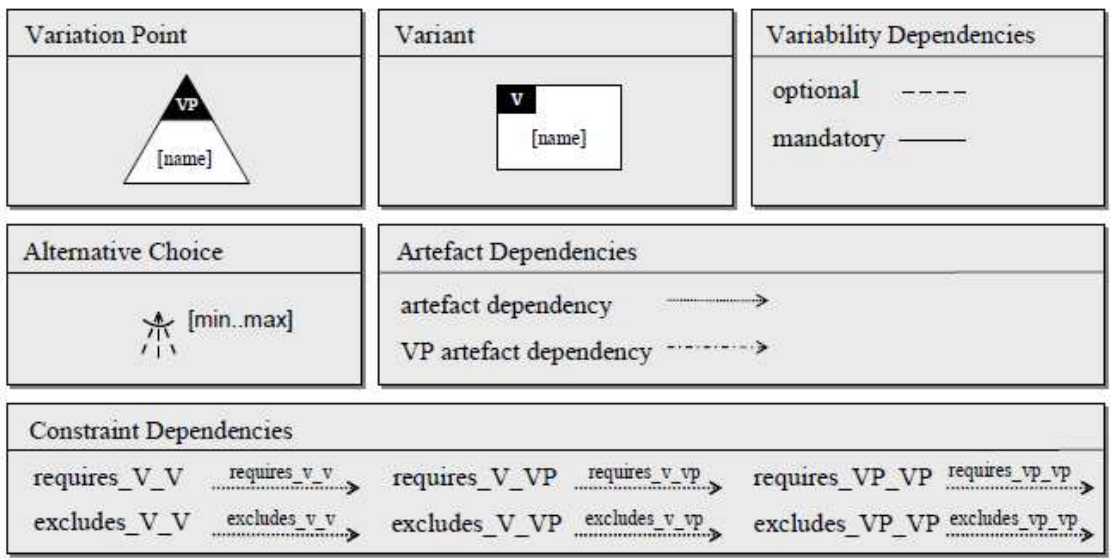

Figure 1. Graphical notation for OVM

In OVM, optional variants may be grouped in *alternative choices*. This group is associated to a cardinality [*min...max*]. Cardinality determines how many Vs may be chosen in an alternative choice, at least min and at most max Vs of the group.

In OVM, constraints between nodes are defined graphically. A constraint may be defined between Vs, VPs, and Vs and VPs. A constraint may be either *excludes* or *requires* constraint. An *excludes* constraint specifies a mutual exclusion, for instance, a variant *excludes* a VP means that if the variant is chosen to a specific product then the VP must not be bound, and vice versa. A *requires* constraint specifies an implication, for instance, a variant requires a VP means that always the variant is part of the product, and the VP must be also part of that product. Figure 2 depicts a simple example of an OVM inspired by the travel agent industry.

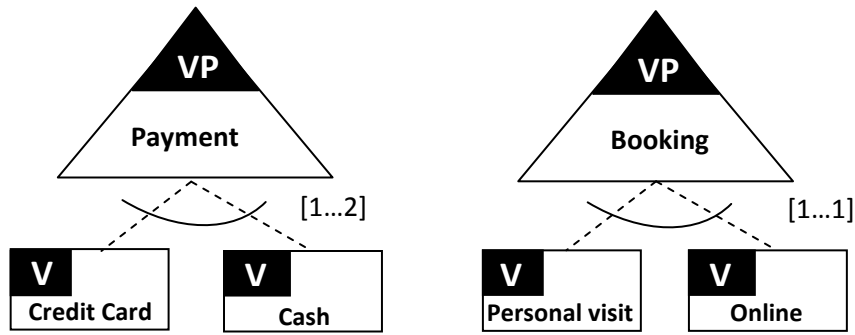

Figure 2. OVM Example: Travel agency industry

## 2.2 Metagraphs

A Metagraph is a graphical structure that represents directed relationships between sets of elements. The theory of Metagraphs is available in [14]. The remainder of this subsection includes a formal representation of a Metagraph, and an example of representing a simple business process as a Metagraph.

The *generating set* of a Metagraph is the set of elements X = {x1, x2, . . . , xn}, which represent variables of interest, and which occur in the edges of the Metagraph. An *edge* e in a Metagraph is a pair e = <$V_e$, $W_e$> ∈ E (where E is the set of edges) consisting of an *in-vertex* $V_e$ ⊂ X and an out-vertex $W_e$ ⊂ X, each of which may contain any number of elements. A Metagraph < X,E> is then a graphical construct specified by its generating set X and a set of edges E defined on the generating set.

An example of a Metagraph is shown in Figure 3. This Metagraph describes the activities and information elements in a travel agency business process. It contains three edges (labeled $e_1$, $e_2$, and $e_3$) that relate various information elements of concern in terms of specific activities. The edge $e_1$ represents the booking activity based upon the customer record; it results in either online booking or personal visit booking. The edge $e_2$ represents the payment activity based on the online booking; it results in credit card payment. The edge $e_3$ represents the payment activity based on the personal visit booking; it results in either credit card payment or cash payment.

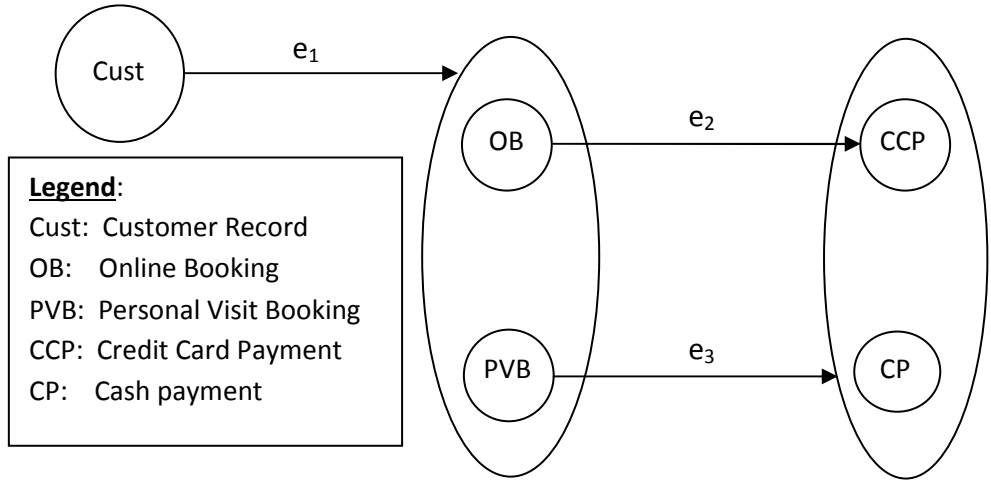

Figure 3. A Travel agency business process Metagraph

### 2.3 AO4BPEL

BPEL (Business Process Execution Language) is an XML-based language that enables task-sharing in a distributed computing environment. Using BPEL [12], a programmer formally describes a business process that will take place across the Web in such a way that any cooperating entity can perform one or more steps in the process the same way.

AO4BPEL [12, 13] is an aspect-oriented extension to BPEL, in which aspects can be (un)plugged into the composition process at runtime. Aspects enable the modularization of concerns that cut across multiple objects. Such concerns are often termed *crosscutting* concerns.

An aspect uses *join points*, point *cuts* and *advices* in the modularization process. A join point is a point during the execution of a program. A Pointcut is a predicate that matches join points. An advice is an action taken by an aspect at a particular join point.

Figure 4, taken from [12], is an example of an aspect. The point cut of this aspect selects the flight search activities in the flight process and in the travel package process. The advice of this aspect contains an activity for starting a timer, which is executed before the join point as well as an activity for stopping a timer, which is executed after the join point.

## 3. Our Approach to SaaS Application Customization

SaaS applications are built following a service-oriented architecture (SOA) as it offers a flexible way for building new composite applications out of existing building blocks [3]. The layers of a SaaS application that need to be customized are: Graphical User Interface (GUI), business process logic (workflow), service, and data [4]. In our proposed approach to customize SaaS applications, we will be concerned only with the workflow and service layers.

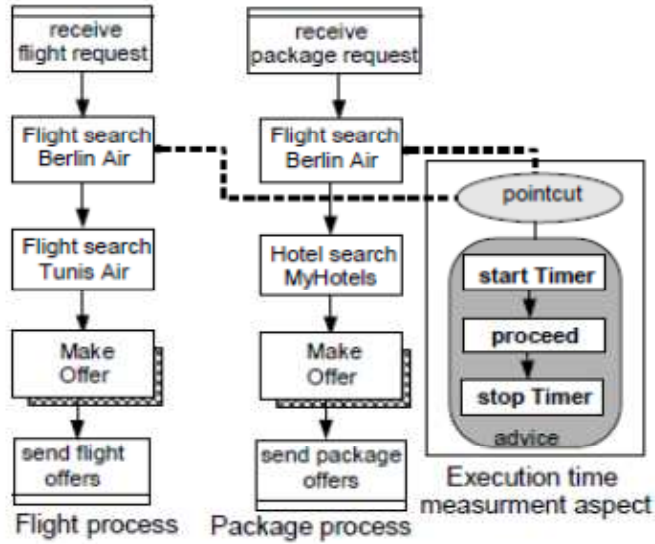

Figure 4. An Aspect Example: Execution time measurement

In our proposed approach, we use OVM to model customizations in workflow and service layers. A Metagraph-based algorithm has been developed to validate tenant-level customizations. AO4BPEL is used to associate and disassociate variations to/from customization points during run-time. Finally, we present a framework for our proposed approach. In the following subsections we demonstrate our approach through a simplified example from the Travel Agency Domain that includes only the booking and payment activities.

## 3.1 Modeling Workflow Customization

In SOA development, a workflow refers to how two or more web services interact [15]. Workflow design often follows the template design pattern. The application designers who design workflows need to define customizable places where the customization is possible. Each customizable place can be replaced by one or more sub-workflow templates. Each sub-workflow contains an atomic or a composite web service, and it can still have customizable places.

The application to model in this paper is taken from the travel agency domain. It contains two activities: booking and payment processes. Figure 5 shows a customizable workflow for this application. It contains two customizable places. Each customizable place in the workflow of our application can be replaced by two sub-workflows. These sub-workflows are given in Figure 6.

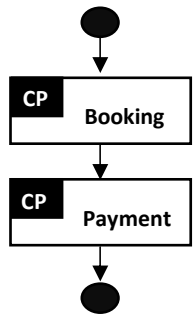

Figure 5. A customizable workflow

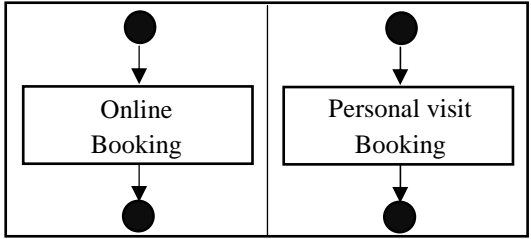
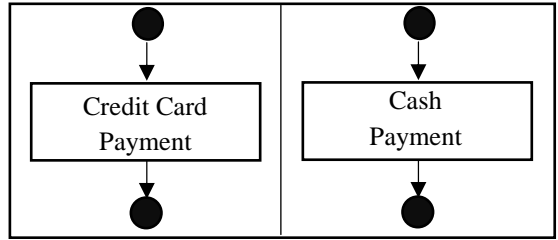

Figure 6.a Sub-workflows for booking      Figure 6.b Sub-workflows for payment

In multi-tenant applications, to provide tenants with understandable customization with constraint dependencies, we model the customizable workflow into OVM. Figure 7 shows the OVM of the customizable workflow for the simplified travel agency application. The customizable places in the workflow are modeled as variation points, and each sub-workflow is modeled as variant.

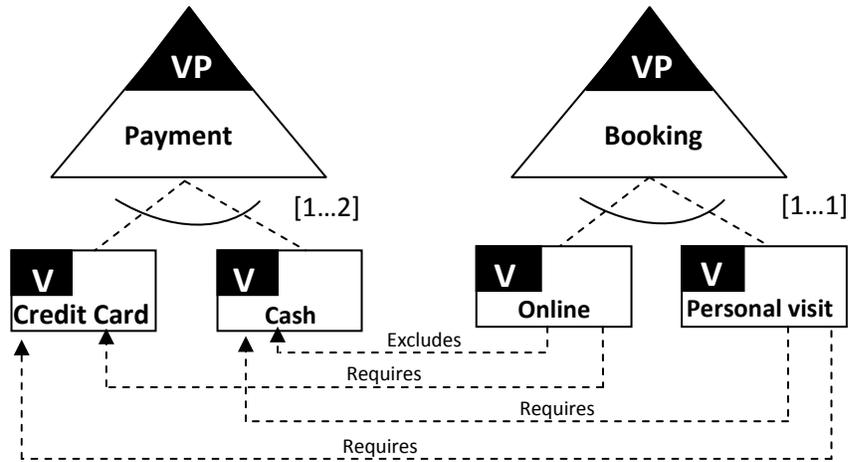

Figure 7. OVM of customizable workflow

## 3.2 Modeling Service Customization

A web service is the main building block to build a SaaS application. A web service customization is the ability to extend, configure, and change web service behavior to achieve particular requirements [16]. To enable SaaS application venders to build highly customizable SaaS applications, the involved web services need to be highly customizable.

To customize a web service, providers need to identify commonalities and variations across the scope of their SaaS application. Identified commonalities are realized as *core services* that exist in all customized applications. Identified variations are realized as *variant services*. Tenants customize web services by selecting one or more of these variation web services.

We model a customizable service into OVM. The customizable service is modeled as a variation point, and each of the core and variant services is modeled as a variant. Figure 8 models a customizable service with one mandatory core service and two optional variant services.

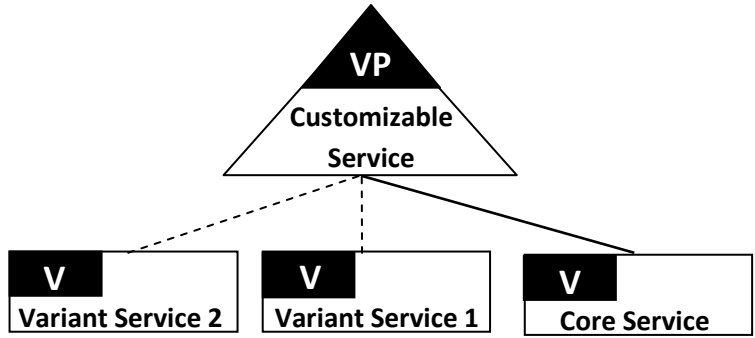

Figure 8: OVM of customizable service

## 3.3 Tenant-Level Customization Validation

In this section we develop a Metagraph algorithm to validate the tenant-level customization that has been modeled as OVM. We start with showing how to map OVMs to Metagraphs, and then we introduce our validation algorithm. Finally, we apply the algorithm on our case study.

Mapping OVMs to Metagraphs:

All variables of interests in OVM (Variation Points and Variants) compose the generating set of the Metagraph that models OVM. Thus, all variation points and variants in OVM are mapped to *vertices* in the Metagraph. The variability and constraint dependencies in OVM are mapped to *edges*, labeled with qualitative attributes, defined on the generating set. Following this mapping procedure, the OVM shown in Figure 7 is mapped to the Metagraph shown in Figure 9.

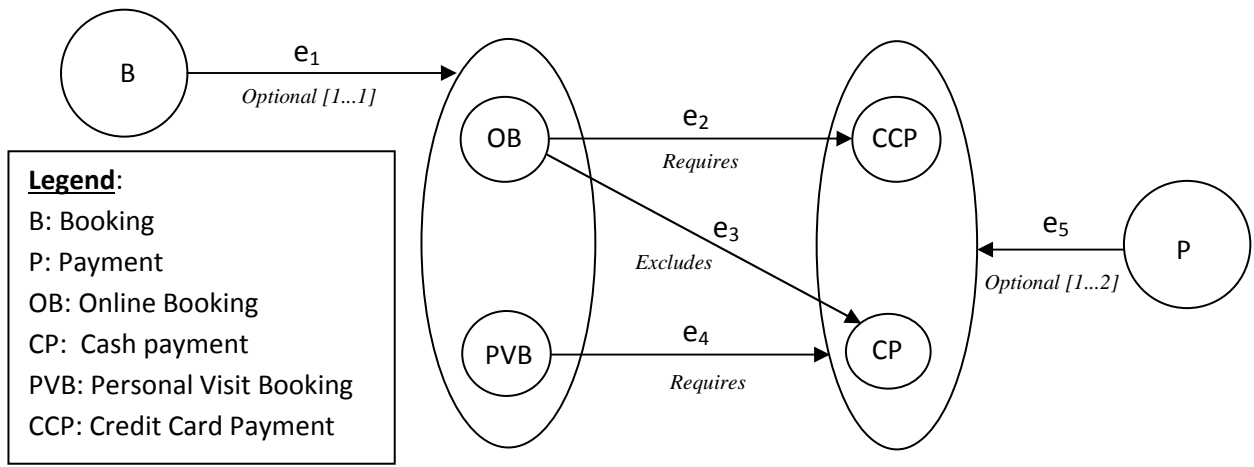

Figure 9. Metagraph representation of the OVM in figure 7

In order to use the above Metagraph in our validation algorithm, we need an algebraic representation of the attributes that are represented visually in the Metagraph. This done by adding each attribute as an additional element to the generating set, and then includes that element in the in-vertex of the edges to which it applies. Applying this procedure to the Metagraph in Figure 9, we get an equivalent Metagraph, shown in Figure 10, in which the qualitative attributes are represented as vertices. The cardinality associated with the alternative choices will be represented as a matrix called the cardinality matrix and shown in Figure 11.

Finally, in order to manipulate the Metagraph, in Figure 10, by our validation algorithm we construct an adjacency matrix representation of the Metagraph. The adjacency matrix A of a Metagraph is a square matrix with one row and one column for each element in the generating set X. The $ij^{th}$ element of A, denoted $a_{ij}$, is a set of triples, one for each edge $e$ connecting $x_i$ to $x_j$. Each triple is of the form <$CI_e$, $CO_e$, $e$>, in which $CI_e$ is the co-input of $x_i$ in $e$ and $CO_e$ is the co-output of $x_j$ in e. Figure 12 shows the adjacency matrix of the Metagraph in Figure 10.

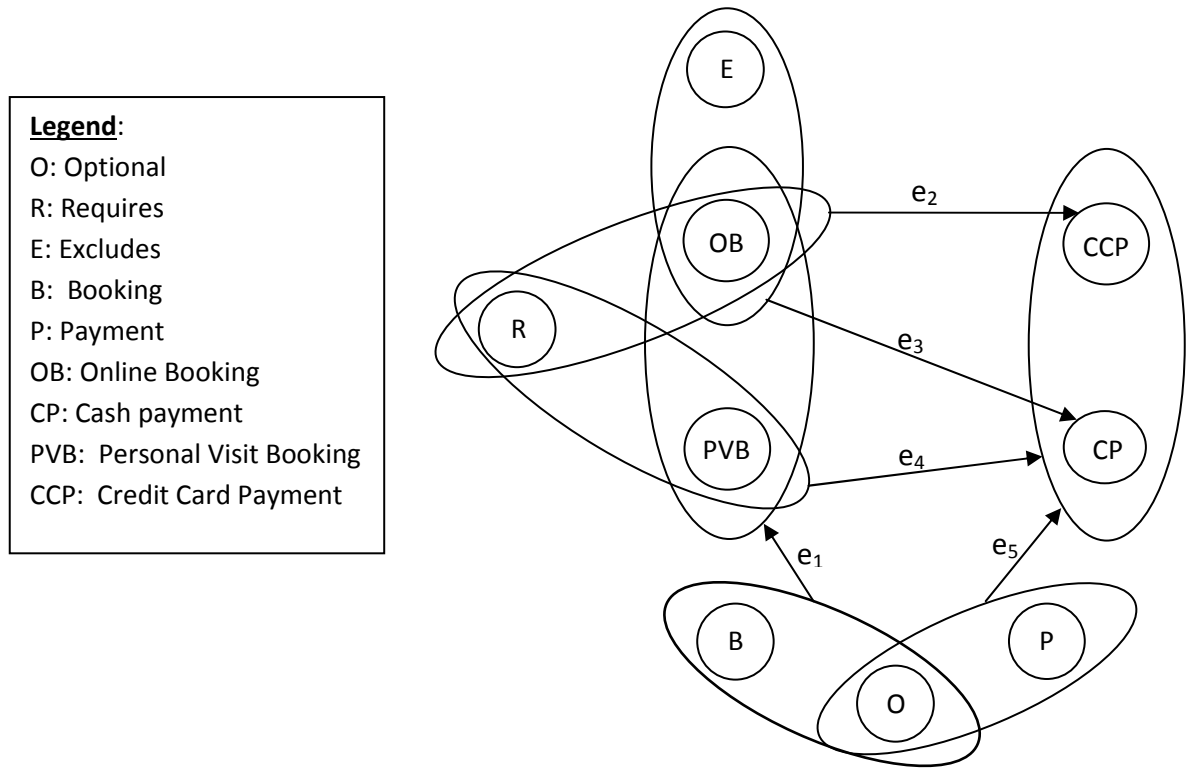

Figure 10. Vertex representation of qualitative attributes

|     | $e_1$ | $e_5$ |
| --- | --- | --- |
| Min | 1 | 1 |
| Max | 1 | 2 |

Figure 11. The cardinality matrix

|     | B | OB | PVB | CCP | CP | P | R | E | O |
|-----|---|----|----|-----|-----|---|---|---|---|
| **B**   | $\varphi$ | <{O},{PVB},$e_1$> | <{O},{OB},$e_1$> | $\varphi$ | $\varphi$ | $\varphi$ | $\varphi$ | $\varphi$ | $\varphi$ |
| **OB**  | $\varphi$ | $\varphi$ | $\varphi$ | <{R}, $\varphi$,$e_2$> | <{E}, $\varphi$,$e_3$> | $\varphi$ | $\varphi$ | $\varphi$ | $\varphi$ |
| **PVB** | $\varphi$ | $\varphi$ | $\varphi$ | $\varphi$ | <{R}, $\varphi$,$e_4$> | $\varphi$ | $\varphi$ | $\varphi$ | $\varphi$ |
| **CCP** | $\varphi$ | $\varphi$ | $\varphi$ | $\varphi$ | $\varphi$ | $\varphi$ | $\varphi$ | $\varphi$ | $\varphi$ |
| **CP**  | $\varphi$ | $\varphi$ | $\varphi$ | $\varphi$ | $\varphi$ | $\varphi$ | $\varphi$ | $\varphi$ | $\varphi$ |
| **P**   | $\varphi$ | $\varphi$ | $\varphi$ | <{O}, {CP},$e_5$> | <{O},{CCP},$e_5$> | $\varphi$ | $\varphi$ | $\varphi$ | $\varphi$ |
| **R**   | $\varphi$ | $\varphi$ | $\varphi$ | <{OB}, $\varphi$,$e_2$> | <{PVB}, $\varphi$,$e_4$> | $\varphi$ | $\varphi$ | $\varphi$ | $\varphi$ |
| **E**   | $\varphi$ | $\varphi$ | $\varphi$ | $\varphi$ | <{OB}, $\varphi$,$e_3$> | $\varphi$ | $\varphi$ | $\varphi$ | $\varphi$ |
| **O**   | $\varphi$ | <{B},{PVB},$e_1$> | <{B},{OB},$e_1$> | $\varphi$ | <{P}, {CCP},$e_5$> | $\varphi$ | $\varphi$ | $\varphi$ | $\varphi$ |

Figure 12. The adjacency matrix of the Metagraph in Figure 10

Customization Validation Algorithm:

Tenant-level customization is performed by assigning variant(s) to each variation point. In our algorithm, the variation points in our application are referred to by CP (Customizable Points) and the variants are referred to by V.

The inputs to our validation algorithm are: (1) M = <X, E>: the Metagraph that models the customizable SaaS application, (2) ICP: Initial Customizable Points, (3) R: the cardinality matrix, and (4) C = {$c_1$, $c_2$, …, $c_n$}: the set of customizations performed by the tenant. Each $c_i$ is represented as an order pair <$cp_i$, $v_i$>, where the $cp_i$ represents a customizable point, and $v_i$ represents the variant assigned to $cp_i$.

The outputs of our validation algorithm are: (1) $M_T$ = <$X_T$, $E_T$>: the Metagraph that models the valid tenant-level customizations, (2) IC: an invalid customization made by the tenant, if found, (3) VF: the validation flag that indicates whether the customization performed by the tenant is valid or not, and (4) CF: the completeness flag that indicates whether the customization performed by the tenant is complete or not.

Figure 13 shows the steps of our validation algorithm. Lines 1-3 construct the adjacency matrix, A, of the Metagraph M, initialize the Metagraph $M_T$ with ICP and all mandatory edges whose in-vertices belong to ICP, and construct the adjacency matrix, $A_T$, of the Metagraph $M_T$.

Lines 4-20 identify the invalid customizations made by the tenant, if any found. Lines 21-25 add the valid customization made by the tenant to the Metagraph $M_T$. Line 26 updates the adjacency matrix, $A_T$, of the Metagraph $M_T$.

A customization instance represented by the pair <$cp_i$, $v_i$> is valid if:
  o All the customizable points involved in required customizations, with customization variant $v_i$, must be selected before selecting $v_i$.
  o The customization variant $v_i$ should not be excluded from the alternative choices for the customizable point $cp_i$.
  o The customizable point $cp_i$ should be selected before selecting the variant $v_i$.

- The customization variant $v_i$ should be one of the customization variants that customize the customizable point $cp_i$.
- Adding the given customization instance should not exceeds the maximum cardinality associated with the alternative choices for the customizable point $cp_i$.

Lines 27-29 make sure that all mandatory customization variants have been associated to the relevant the customization points in $X_T$. Lines 30-32 make sure that the minimum cardinality associated to any alternative choice group is achieved. Finally, Line 33 returns two flags: the flag CF indicates that the customization made by the tenant is complete. The flag VF indicates that all customization instances made by the tenant are valid.

```
Begin
1: Construct the adjacency matrix, A, of the Metagraph M.
2: Initialize M_T: (X_T = ICP and E_T = All mandatory edges whose in-vertices belong to ICP)
3: Construct the adjacency matrix, A_T of the Metagraph M_T.
4: for each element c_i in the customization set C. (c_i = <cp_i, v_i>)
5:   for each non-empty element a[v_i][j] in the row v_i of the matrix A
6:     for each triple <CI_e, CO_e, e> in a[v_i][j]
7:       if the requires-vertex ∈ CI_e and x_j ∉ X_T, where x_j refers to the vertex corresponding to the column j.
8:         CF = incomplete, IC = c_i , VF = invalid, and return
9:   for each non-empty element a[j][v_i] in the column v_i of the matrix A
10:    for each triple <CI_e, CO_e, e> in a[j][v_i],
11:      if the excludes-vertex ∈ CI_e and x_j ∈ X_T, where x_j refers to the vertex corresponding to the row j.
12:        CF = incomplete, IC = c_i , VF = invalid, and return
13:  if cp_i ∉ X_T
14     CF = incomplete, IC = c_i , VF = invalid, and return
15:  if the element a[cp_i][v_i] in A is empty
16:     CF = incomplete, IC = c_i , VF = invalid, and return
17:  e′ = e, where e is the edge in the triple ⟨CI_e, CO_e, e⟩ ∈ a[cpi][vi] in A, such that the optional-vertex ∈ CI_e
18:  m =  The union of the co-outputs of all triples ⟨CI_e, CO_e, e⟩ in the row cp_i of A_T, such that e = e′
19:  if m = R[max][e′]
20:     CF = incomplete,  IC = c_i , VF = invalid, and return
21:  X_T = X_T  ∪{v_i}
22:  If  e′ ∈ E_T and  e′ = ⟨V_e′, W_e′⟩
23:     W_e′ = W_e′ ∪ {vi}
24:  else
25:     E_T = E_T ∪ {e′}, where e′ = ⟨V_e′, {vi}⟩
26: Update A_T
27: for each triple ⟨CI_e, CO_e, e⟩ in the mandatory row in the matrix A
28:    if CI_e ⊆X_T and CO_e ⊈ X_T
29:      CF = incomplete, VF = invalid and return
30: for each triple ⟨CI_e, CO_e, e⟩ in the optional row in the matrix A_T
31:    if |CO_e| ≤ R[min][e],
32:      CF = incomplete, VF = invalid and return
33: CF = complete, VF = valid and return
End
```

Figure 13. Customization validation algorithm

## 3.4 Runtime Customization using AO4BPEL

Each *customization point* in a SaaS application has one or more *customization variants*. Tenants customize the SaaS application by selecting a customization variant and associate it to the customization point. Tenants may disassociate a customization variant from a customization point and replace it with another customization variant. In our approach, we use AO4BPEL to enable modular implementation of these customizations and dynamic adaptation during run-time.

By using AO4BPEL, customization variants can be composed to or decomposed from their customization points during run-time without stopping, rebinding, recompiling, or even restarting the applications. In our approach, every customization point is designed as a BPEL process, and each customization variant is designed as an aspect. All BPEL processes and the all aspects are developed by the *SaaS application developer*.

For the application presented in this paper, we need to design two BPEL processes for the booking and payment customization points. In addition, we need to design four aspects for the customization variants (online booking, personal booking, cash payment, and credit card payment). Based on the customization made by the tenant, the AO4BPEL engine dynamically weaves the selected aspects (customization variants) with their corresponding processes (customization points).

All the files related to the BPEL processes, the aspects, the weaving process, and the involved web services (source code, WSDLs, and file descriptors) are stored in three stores: *Process Store, Aspect Store, and Service Store*.

## 3.5 The Framework of our Customization Approach

This subsection summarizes the activities involved in customizing a SaaS application. Figure 14 describes the customization scenarios based on our customization approach. The *Customization-Validation* unit, which is an implementation of our customization validation algorithm, takes as inputs: (1) the SaaS customization data prepared by the SaaS application developer and stored in the *SaaS-Customization-Data* database, (2) the tenants' customization set prepared by the tenants' administrators, and (3) the previous validated customizations made by the same tenant's administrator and stored in the *Validated-Customization-Data* database. The output of the validation process will be stored in the *Validated-Customization-Data* database.

The SaaS application developer implements all valid customizable points as processes and stores them the *Process Store* unit. The developer also implements all valid customization variants as aspects and stores them in the *Aspect Store* unit. All web services that are related to the developed processes and aspects are stored in the *Service Store* unit.

The *Validation UI* unit enables the tenants' administrators to define their customization sets, and send a request to the *Customization-Validation* unit to validate these sets. It also enables the tenants' administrators to make the necessary actions if the customization set is incomplete, or if any customization instance is invalid.

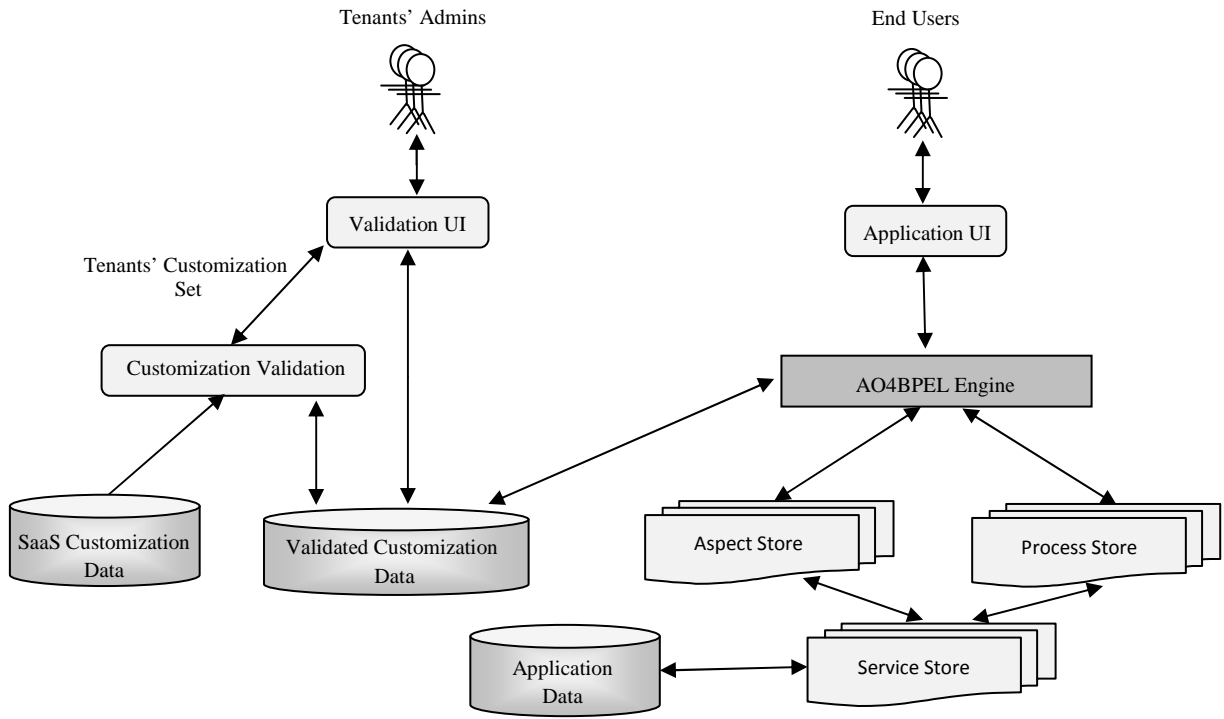

Figure 14. The Framework of our Proposed Approach

Based on a request from an end user, the AO4BPEL engine retrieves the relevant validation customization data for its tenant from the *Validated-Customization-Data* database, and then waves the corresponding process and aspects, and finally, the engine loads the waved process to perform the user request.

## 4. Comparison with Related Work

Customization could be performed in two different approaches: *Source-Code Based* or *Composition Based*. In the source-code based customization, SaaS applications are customized by adding and integrating new source code with the customizable application. Examples of research papers that follow this approach include [2, 5, 7, 17, 18].

The source-code based customization approach although it provides tenants with flexibility in the customization process, it suffers from some drawbacks including: (1) The tenant administrator should be aware of the implementation details of the SaaS application, (2) Allowing the tenant administrator to add and integrate source code may violate the security regulations of the application, and (3) The source code added by the tenant should be revised when the SaaS application developer upgrade the application.

In the composition based approach, SaaS applications are customized by composing variant components, selected from a provided set of components, to the customizable application. In addition to our proposed approach in this paper, examples of other research papers that follow this approach include [3, 8, 10, 15,19, 20].

As mentioned earlier in the introduction, there are four key concerns that need to be addressed in the SaaS application customization: first, how to model the customization points and variations, second, how to describe the relationships among variations, third, how to validate customizations performed by tenants, and fourth, how to associate and disassociate variations to/from customization points during run-time.

In our approach we dealt with all the above mentioned concerns by providing the tenant with simple and understandable customization model, developing a customization validation algorithm, and making use of the aspect-oriented approach to handle the runtime customization. On the other hand, many of the other related work partially addressed these concerns as discussed below.

In [10], the authors dealt with three concerns by using Metagraphs to model customization points, variants, and their relationships. They also proposed a validation algorithm to validate customizations made by tenants. However, the proposed model dealt only with the "*requires*" relationship. On the other hand, Metagraph-based modeling is very difficult for tenants to understand, especially in large applications with thousands of variants.

The authors in [4] handled only one concern that is modeling customization points, and variants using ontology-based intelligent customization framework. In [15], the authors also dealt with only one concern that is modeling customization points, and variants using OVM. They also guided the tenant through the customization process.

Finally, in [3], the authors dealt with three concerns by using OVM to model customization points, variants, and their relationships. To avoid unpredictable customizations, tenants are guided through the customization process. However, in large applications with thousands of variants, the guiding process they proposed will not be very helpful for the tenant.

## 5. Conclusion

This paper discussed the concerns that need to be addressed to customize SaaS applications, and then proposed an aspect-oriented approach to address these concerns. OVM is used to model customizations in workflow and service layers. A Metagraph-based algorithm has been developed to validate tenant-level customizations. AO4BPEL is used to associate and disassociate variations to/from customization points during run-time. We demonstrated our approach through a simplified example from the Travel Agency Domain. Finally, we compare our proposed approach with others in related work.